\documentclass[12pt,aps,epsfig,amssymb,tighten]{revtex4}
\usepackage{epsfig}
\begin{document}
\title{Equilibrium statistics of an inelastically bouncing ball,\\
subject to gravity and a random force}
\author{
Theodore W. Burkhardt$^1$ and Stanislav N. Kotsev$^{1,2}$}
\affiliation{$^1$Department of Physics, Temple University, Philadelphia, PA 19122\\
$^2$Department of Chemistry, Rice University, Houston, TX 77005}
\date{\today}

\begin{abstract}
We consider a particle moving on the half line $x>0$ and subject to a constant force
in the $-x$ direction plus a
delta-correlated random force. At $x=0$ the particle is
reflected inelastically. The velocities just after and before reflection satisfy
$v_f=-rv_i$, where $r$ is the coefficient of restitution. This simple model is of
interest in connection with studies of driven granular matter in a gravitational
field. With an exact analytical approach and simulations we study the steady state
distribution function $P(x,v)$.
\end{abstract}
\maketitle

\clearpage
\section{Introduction}
\label{sec:intro} Consider a particle moving on the half line $x>0$ and subject to a
constant force in the $-x$ direction, such as gravity, and an additional
delta-correlated random force with zero mean. The Newtonian equation of motion has
the form
\begin{equation}{d^2x\over
dt^2}=-g+\eta(t)\thinspace,\quad\langle \eta(t)\eta(t')\rangle=
2\Lambda\delta(t-t')\;,\label{eqmo}
\end{equation}
where $g$ is a positive constant. With no loss of generality we set $g=1$,
$\Lambda=1$ throughout this paper, since this can be achieved by rescaling
\cite{rescale} of $x$ and $t$. At the boundary $x=0$ the particle is reflected
inelastically. The velocities just after and before reflection satisfy
\begin{equation}
v_f=-rv_i\thinspace,\label{restitution}
\end{equation}
where $r$ is the coefficient of restitution. The simple model defined by Eqs.
(\ref{eqmo}) and (\ref{restitution}) is of interest in connection with driven
granular gases, see, e.g., \cite{mss}. Cornell, Swift, and Bray \cite{csb}
originally proposed the model and studied the case $g=0$.

According to Eq. (\ref{eqmo}) the mean energy $E(t)=\left\langle {1\over 2}v(t)^2
+x(t)\right\rangle$ of the randomly accelerated particle at time $t$ increases as
$E(t)=E(0)+t$ between boundary collisions but decreases, for $r<1$, in the
collisions. Eventually an equilibrium is reached. In this paper we study the
equilibrium distribution function $P(x,v)$ with an exact analytical approach and
simulations. In related earlier work, but without gravity, we have considered (i)
the equilibrium of a randomly accelerated particle moving on the finite line between
two inelastically reflecting boundaries \cite{bfg,bk}, and (ii) non-equilibrium
properties of a randomly accelerated particle on the half line \cite{twb93,twb2000}
and on the finite line \cite{kb} for absorbing, partially absorbing, and inelastic
boundary conditions .

The equilibrium distribution function $P(x,v)$ of a particle, which moves according
to Eq. (\ref{eqmo}) for $x>0$ and is reflected inelastically at $x=0$, satisfies the
steady-state Fokker-Planck equation \cite{fpeq}
\begin{equation}
\left(v{\partial\over\partial x}-{\partial\over
\partial v}-{\partial^2\over
\partial v^2}\right)P(x,v)=0\thinspace,\label{fp}
\end{equation}
with the boundary condition
\begin{equation}
P(0,-v)=r^2P(0,rv)\;\;,\quad v>0\;.\label{bc}
\end{equation}
The factors of $r$ in Eq. (\ref{bc}) take the change in velocity (\ref{restitution})
into account and ensure that the incident and reflected probability currents at the
boundary have equal magnitude
\begin{equation} I= \int_0^\infty
dv\thinspace vP(0,-v)=\int_0^\infty dv\thinspace
vP(0,v)\thinspace.\label{collrate}
\end{equation}
Here we have assumed the standard normalization,
\begin{equation} \int_0^\infty dx\int_{-\infty}^\infty dv\thinspace
P(x,v)=1\;,\label{normalization}
\end{equation}
of the distribution function.

The main goal of the paper is to solve the steady-state Fokker-Planck equation
(\ref{fp}) with boundary condition (\ref{bc}).  In Section II we derive an integral
equation for the distribution function $P(0,v)$ at the boundary and show how to
obtain $P(x,v)$ from $P(0,v)$ by integration. In Section III the asymptotic form of
$P(0,v)$ for large and small $v$ is determined, for arbitrary $r$, from the integral
equation. For two particular values, $r={1\over 2}$ and ${1\over 3}$, the integral
equation is solved exactly and the corresponding $P(x,v)$ obtained by integration.
We have also carried out numerical simulations of the randomly accelerated
accelerated particle in a gravitational field. The algorithm is described in Section
IV, and the results of the simulations are compared with our analytic predictions.
Section V contains closing remarks.

\section{Integral Equation for $P(x,v)$}
The Fokker-Planck equation (\ref{fp}) is readily solved by separation of variables.
The most general solution which vanishes for $x\to\infty$ and $v\to\pm\infty$ is
given by \cite{twb93}
\begin{equation}
P(x,v)=e^{-v/2}\int_0^\infty dF\thinspace
w(F)e^{-Fx}\psi_{1/4,F}(-v)\;,\label{fpsol}
\end{equation}
where $w(F)$ is an arbitrary function, and $\psi_{s,F}(v)$, with $F>0$, is the
solution of the ordinary differential equation
\begin{equation}
\left(-{d^2\over dv^2}+Fv+s\right)\psi_{s,F}(v)=0\label{diffeq}
\end{equation}
which vanishes for $v\to\pm\infty$. This solution is given by
\begin{equation}
\psi_{s,F}(v)=F^{-1/6}{\rm Ai}\left(F^{1/3}v+F^{-2/3}s\right)\;, \label{psi}
\end{equation}
where ${\rm Ai}(z)$ is the standard Airy function \cite{as}. Normalized as in Eq.
(\ref{psi}), the $\psi_{s,F}(v)$ have the orthogonality and closure properties
\cite{twb93}
\begin{equation}
\int_{-\infty}^\infty dv\thinspace v\psi_{s,F}(\pm v)\psi_{s,G}(-v)=
\left\{\begin{array}{l}0\;,\nonumber\\
\delta(F-G)\;,\end{array}\right.\begin{array}{l}{\rm upper\ sign}\nonumber\\
{\rm lower\ sign}\nonumber\end{array}\label{orthonorm1}
\end{equation}
\begin{equation} \int_0^\infty dF\thinspace
\left[\;\psi_{s,F}(-v)\psi_{s,F}(-u)-\psi_{s,F}(v)\psi_{s,F}(u)\;\right]=
v^{-1}\delta(v-u)\;.\label{closure}
\end{equation}

Making use of Eqs. (\ref{fpsol}), (\ref{psi}), and the relation
$\int_{-\infty}^\infty dt \thinspace e^{-at}{\rm Ai}(-t)=\exp\left(a^3/3\right)$ for
$a>0$, one finds that the probability density $P(x)$ for the position of the
particle is given by
\begin{equation}
P(x)=\int_{-\infty}^\infty dv\thinspace P(x,v)=\int_0^\infty dF\thinspace
F^{-1/2}w(F)\exp\left[-(12F)^{-1}-Fx\right]\;,\label{P(x)1}
\end{equation}
and that the normalization condition (\ref{normalization}) is satisfied if
\begin{equation} \int_0^\infty dx\thinspace P(x)=\int_0^\infty dF\thinspace
F^{-3/2}w(F)\exp\left[-(12F)^{-1}\right]=1\;.\label{normalization2}
\end{equation}

To invert Eq. (\ref{fpsol}), it is convenient to use the auxiliary functions
\cite{twb93}
\begin{equation}
\phi_{s,F}(v)=\psi_{s,F}(v)-\int_0^\infty{dG\over 2\pi}\;{\exp\left[-{2\over
3}s^{3/2}\left(F^{-1}+G^{-1}\right)\right]\over F+G}\;\psi_{s,G}(-v)\;,\label{phi}
\end{equation}
which vanish for $v>0$ and have the biorthogonality property
\begin{equation}
\int_0^\infty dv\thinspace v\phi_{s,F}(-v)\psi_{s,G}(-v)=\delta(F-G)\label{biortho}
\end{equation}
on the semi-infinite interval $v>0$. Functions of this type were first constructed
by Marshall and Watson \cite{mw} and exploited in \cite{twb93,mw,kainz} in
calculating first passage properties.

Using property (\ref{biortho}) to solve Eq. (\ref{fpsol}) for $w(F)$, we obtain
\begin{equation}
w(F)=\int_0^\infty dv\;v\phi_{1/4,F}(-v)e^{v/2}P(0,v)\;.\label{weightfunc}
\end{equation}
Substituting this result back in Eq. (\ref{fpsol}) and imposing the boundary
condition (\ref{bc}) leads to
\begin{eqnarray}
&&r^2P(0,rv)=\int_0^\infty du\thinspace u G_{g}(v,u)P(0,u)\;,\label{inteq}\\
&&G_g(v,u)=e^{(v+u)/2}\int_0^\infty dF\thinspace
\psi_{1/4,F}(v)\phi_{1/4,F}(-u)\;.\label{kernel1}
\end{eqnarray}
Equations (\ref{fpsol}) and (\ref{weightfunc})-(\ref{kernel1}) provide us with a
strategy for calculating $P(x,v)$. First $P(0,v)$ is determined by solving the
integral equation (\ref{inteq}). Then $w(F)$ and $P(x,v)$ are obtained by evaluating
the integrals in Eqs. (\ref{weightfunc}) and (\ref{fpsol}), respectively.

The kernel in integral equation (\ref{inteq}) has a simple physical interpretation.
The quantity $vG_g(v,u)dv$ is the probability that a particle moving according to
Eq. (\ref{eqmo}) with initial position $x=0$ and initial velocity $u>0$ arrives with
speed between $v$ and $v+dv$ on its first return to $x=0$. Integral equation
(\ref{inteq}) for the equilibrium distribution follows immediately from this
interpretation. For consistency with the probabilistic interpretation, $G_g(v,u)$
should be symmetric in $v$ and $u$ and normalized so that
\begin{equation}
\int_0^\infty dv\thinspace v G_g(v,u)=1\;.\label{normG}
\end{equation}
That the expression for $G_g(v,u)$ in Eq. (\ref{kernel1}) satisfies both these
conditions can be shown with the help of Eqs. (\ref{psi}), (\ref{closure}), and
(\ref{phi}).

The integral representation of $G_g(v,u)$ in Eqs. (\ref{kernel1}) is useful because
of the simple orthogonality and closure properties of the basis functions. Two other
useful representations,
\begin{eqnarray}
G_g(v,u)&=&{e^{(v+u)/2}\over \pi^2 vu}\int_0^\infty
d\gamma\thinspace\gamma\thinspace{\sinh(\pi\gamma)\over
\cosh(\pi\gamma/3)}\thinspace
K_{i\gamma}(v)K_{i\gamma}(u)\label{kernel2}\\
G_g(v,u)&=&{3\over 2\pi}(vu)^{1/2}e^{(v+u)/2}\int_0^1 dy\left\{\thinspace
\exp\left[-(v^2-vu+u^2+3vuy^2)^{1/2}\right]\right.\nonumber\\
&\times&\left.\left[(v^2-vu+u^2+3vuy^2)^{-3/2}+(v^2-vu+u^2+3vuy^2)^{-1}
\right]\right\}\;,\label{kernel3}
\end{eqnarray}
follow from a classic paper of McKean \cite{mck}, as shown in Appendix A. In Eq.
(\ref{kernel2}), $K_\nu(z)$ is the standard modified Bessel function \cite{as,gr}.
Expressions (\ref{kernel2}) and (\ref{kernel3}) are particularly convenient for
analytical (see Appendix B) and numerical calculations, respectively.

\section{Solution of Integral Equation (\ref{inteq})}
\subsection{Asymptotic form of $P(0,v)$ for small $v$}

First we derive the asymptotic form of $P(0,v)$ for small and large $v$ from
integral equation (\ref{inteq}). Using representation (\ref{kernel3}) for the
kernel, one finds that the ansatz
\begin{equation}
P(0,v)\sim v^{-\beta(r)}\;,\quad v\searrow 0\;\label{asym1}
\end{equation}
is consistent with the integral equation if the exponent $\beta(r)$ satisfies
\begin{equation}
r=\left[2\sin\left({2\beta+1\over
6}\thinspace\pi\right)\right]^{1/(\beta-2)}\;.\label{asym2}
\end{equation}

The dependence of $\beta$ on the coefficient of restitution $r$ is shown in Fig. 1.
As $r$ decreases from 1 to 0, $\beta(r)$ increases monotonically from 0 to 5/2. The
probability density for finding the particle at the boundary with velocity $v$
diverges as $v$ approaches zero, and the greater the inelasticity, the stronger the
divergence.

The exponent $\beta$ has the value 2 at
\begin{equation}
r_c=e^{-\pi/\sqrt 3}=0.163\dots\label{rc}
\end{equation}
For $r<r_c$, $\beta(r)>2$, so that $P(0,v)$ diverges more strongly than $v^{-2}$ in
the limit $v\searrow 0$. Thus, for $r<r_c$ the mean collision rate, defined by the
integral (\ref{collrate}), diverges at the lower limit, i.e., the randomly
accelerated inelastic particle makes an infinite number of collisions in a finite
time. Cornell et al. \cite{csb} argued that even in the absence of a gravitational
field this is the case.

The small $v$ behavior (\ref{asym1}), (\ref{asym2}) is precisely the same as for a
randomly accelerated particle moving on the finite line $0<x<1$ with inelastic
collisions at $x=0$ and $1$ and with no gravitational field \cite{bfg,bk}. In both
cases the divergent behavior for $v\searrow 0$ comes from the high rate of low
velocity inelastic collisions with one of the walls. For short times the diffusive
change in velocity due to the random force, which grows as $t^{1/2}$, exceeds the
change due to the gravitational force, which is proportional $t$. Thus, the random
force is primarily responsible for the rapid return to the boundary and determines
the small $v$ behavior.

In both cases - the randomly accelerated particle on the half line with a
gravitational field considered here, and on the finite line with no gravitational
field considered in \cite{bk} - the distribution function $P(x,v)$ remains smooth,
extended, and normalizable for $r<r_c$. In neither case does it collapse on the
boundary. Thus, the infinite equilibrium collision rate does not lead to
localization of the particle at the boundary.

\subsection{Asymptotic form of $P(0,v)$ for large $v$}
For large $v$ one finds that the ansatz
\begin{equation}
P(0,v)\sim v^{-m(r)}e^{-q(r) v}\;,\quad v\to\infty\;.\label{asym3}
\end{equation}
is consistent with integral equation (\ref{inteq}) with kernel (\ref{kernel3}) if
the quantities $m(r)$ and $q(r)$ are chosen as follows: For ${1\over 2}\leq r<1$,
\begin{equation}
m(r)=1-{\ln r\over\ln[x(r)/r]}\;,\quad q(r)={3\over 2}\thinspace{1-r\over
1-r+r^2}\;,\label{asym4}
\end{equation}
where
\begin{equation}
x(r)={1\over 2}+{\sqrt 3\over 2}\thinspace{{1\over 2}-q\over \left[1-\left({1\over
2}-q\right)^2\right]^{1/2}}\;,\label{asym5}
\end{equation}
and for $r<{1\over 2}$,
\begin{equation}
m(r)={3\over 2}\;,\quad q(r)={1\over 2r}\;.\label{asym6}
\end{equation}

The functions $m(r)$ and $q(r)$ are shown in Figs. 1 and 2, respectively. As $r$
decreases from 1 to ${1\over 2}$, $m(r)$ increases from 0 to 1 and then changes
discontinuously to ${3\over 2}$ for $r<{1\over 2}$. As $r$ decreases from 1 to 0,
$q(r)$ increases monotonically from 0 to $\infty$. The functional form of $q(r)$
changes at $r={1\over 2}$, in accordance with Eqs. (\ref{asym4}) and (\ref{asym6}).
Both $q$ and $dq/dr$ are continuous at $r={1/2}$, but $d^2q/dr^2$ is discontinuous.

The functions $m(r)$ and $q(r)$ are both non-analytic at $r={1\over 2}$. Since they
only characterize $P(0,v)$ in the large $v$ limit (see Eq. (\ref{asym3})), it does
not follow that $P(x,v)$ is non-analytic in $r$ for fixed finite $x$ and $v$. We
have not found any indication of a phase transition, i.e., a non-analyticity of the
distribution function $P(x,v)$, as $r$ is varied with $x$ and $v$ fixed.

\subsection{Exact $P(0,v)$ for $r={1\over 2}$ and ${1\over 3}$}
The curves for $\beta(r)$ and $m(r)$ in Fig. 1 cross at $r={1\over 2}$ and ${1\over
3}$. For these two values of $r$, the exponents in the asymptotic forms
(\ref{asym1}) and (\ref{asym3}) for small and large $v$ are the same, with
$\beta({1\over 2})=m({1\over 2})=1$ and $\beta({1\over 3})=m({1\over 3})={3\over
2}$. In fact, the functions
\begin{equation}
P(0,v)=\left\{\begin{array}{l}A_{1/2}\; v^{-1}e^{-v}\;,\\
A_{1/3}\; v^{-3/2}e^{-3v/2}\;,
\end{array}\right.\quad\begin{array}{l}r={1\over2}\;,\\r={1\over
3}\;,\end{array}\label{P(0,v)}
\end{equation}
turn out to be exact solutions of integral equation (\ref{inteq}) for all
$0<v<\infty$, as we discovered numerically and prove analytically in Appendix B. The
quantities $A_{1/2}$ and $A_{1/3}$ in Eq. (\ref{P(0,v)}) are normalization
constants, which we now determine.

Substituting Eq. (\ref{P(0,v)}) in Eq. (\ref{weightfunc}) and making use of Eqs.
(\ref{psi}) and (\ref{phi}), we find that the corresponding $w(F)$ are given by
\begin{equation}
w(F)=\left\{\begin{array}{l}{1\over 2}A_{1/2}\; F^{-1/2}
\exp\left[-(12F)^{-1}\right]\;,\\
\left({2\over 27\pi}\right)^{1/2}A_{1/3}\; F^{-1/2}K_{1/6}\left((12F)^{-1}\right)\;,
\end{array}\right.\quad\begin{array}{l}r={1\over2}\;,\\r={1\over 3}\;,\end{array}
\label{weightfunc2}\end{equation} where $K_\nu(z)$, as in Eq. (\ref{kernel2}), is a
modified Bessel function. Together with Eq. (\ref{normalization2}), these $w(F)$
imply
\begin{equation}
A_{1/2}={1\over 3}\;,\quad A_{1/3}=\left(27\over 32\pi\right)^{1/2}\label{constants}
\end{equation}
for the normalization constants in Eqs. (\ref{P(0,v)}) and (\ref{weightfunc2}).

Combining Eqs. (\ref{P(x)1}), (\ref{weightfunc2}), and (\ref{constants}), one
obtains relatively simple expressions
\begin{equation}
P(x)=\left\{\begin{array}{l}{1\over 3}K_0(\sqrt{2x/3})\;,\\{1\over
4\pi}\int_0^\infty dF\thinspace
F^{-1}K_{1/6}\left((12F)^{-1}\right)\exp\left[-(12F)^{-1}-Fx\right]\;,\\
\end{array}\right.\quad\begin{array}{l}r={1\over2}\;,\\r={1\over 3}\;,\end{array}
\label{P(x)2}\end{equation} for the probability distribution of position. In the
limit $x\to 0$, $P(x)$ diverges as $\log x$ and $x^{-1/6}$ for $r={1\over 2}$ and
${1\over 3}$, respectively, and as $x^{(1-\beta)/3}$ for general $r$, as in Ref.
\cite{bk}. In the limit $x\to\infty$, $P(x)$ decays as $x^{-1/4}\exp(-\sqrt{2x/3})$
and $x^{-1/2}\exp(-\sqrt{2x/3})$ for $r={1\over 2}$ and ${1\over 3}$, respectively.

The distribution function $P(x,v)$ may be calculated numerically for $r={1\over 2}$
and ${1\over 3}$ by combining Eqs. (\ref{fpsol}), (\ref{psi}), (\ref{weightfunc2}),
and (\ref{constants}) and carrying out the integration over $F$ numerically. The $v$
dependence of $P(x,v)$  for $r={1\over 2}$ and several values of $x$ is shown in
Fig. 3. For $r={1\over 3}$ the curves are qualitatively similar. In both cases the
curve for $x=0$ diverges at $v=0$, in accordance with Eqs. (\ref{asym1}) and
(\ref{asym2}). For $x>0$ the divergence is replaced by a smooth peak, which becomes
lower and broader as $x$ increases.

For large $x$ the height of the peak in Fig. 3 varies as $P(x,0)\sim
x^{-1/2}\exp(-\sqrt{2x/3})$ for $r={1\over 2}$ and as $P(x,0)\sim
x^{-3/4}\exp(-\sqrt{2x/3})$ for $r={1\over 3}$. For small $x$, $P(x,0)\sim x^{-1/3}$
and $x^{-1/2}$ for $r={1\over 2}$ and ${1\over 3}$, respectively, and $P(x,0)\sim
x^{-\beta/3}$ for general $r$, as in Ref. \cite{bk}.

The curves in Fig. 3 are characterized by the moments
\begin{equation}
\langle v^n\rangle_x=P(x)^{-1}\int_{-\infty}^\infty dv\;v^nP(x,v)\;,\label{moments}
\end{equation}
where the normalizing denominator $P(x)$ is defined by Eq. (\ref{P(x)1}). Taking the
derivative of Eq. (\ref{moments}) with respect to $x$, making use of the
Fokker-Planck equation (\ref{fp}), and integrating partially with respect to $v$,
one finds that
\begin{eqnarray}
&&\langle v\rangle_x=0\;,\label{moment1}\\
&&\langle v^2\rangle_x=P(x)^{-1}\int_x^\infty dx\;P(x)\;.\label{moment2}
\end{eqnarray}
According to Eqs. (\ref {collrate}), (\ref{moments}), and (\ref{moment1}) the net
probability current at a distance $x$ from the boundary vanishes, as expected in
equilibrium. Inserting the asymptotic form of $P(x)$, given just below Eq.
(\ref{P(x)2}), in Eq. (\ref{moment2}), one finds that the root-mean-square width
$\langle v^2\rangle_x^{1/2}$ of the peak in Fig. 3 increases as $x^{1/4}$ in the
large $x$ limit, for both $r={1\over 2}$ and ${1\over 3}$.

\section{Simulations}
\subsection{Algorithm}
In our simulations the motion of the particle is governed by
the difference equations
\begin{eqnarray}
x_{n+1}&=&x_n+v_n\Delta_{n+1}-{1\over 2}\;\Delta_{n+1}^2+\left({1\over
6}\;\Delta_{n+1}^3\right)^{1/2}(s_{n+1}+\sqrt{3}\thinspace r_{n+1})\;,\label{xstep}\\
v_{n+1}&=&v_n-\Delta_{n+1}+(2\Delta_{n+1})^{1/2}\thinspace r_{n+1}\;.\label{vstep}
\end{eqnarray}
Here $x_n$ and $v_n$ are the position and velocity at time $t_n$, and
$\Delta_{n+1}=t_{n+1}-t_n$. The quantities $r_n$ and $s_n$ are independent Gaussian
random numbers with $\langle r_n\rangle=\langle s_n\rangle=0$ and $\langle
r_n^2\rangle=\langle s_n^2\rangle=1$. This is the same as the algorithm in Refs.
\cite{bibu,kb}, except that a gravitational field $g=1$ has been included.

The algorithm (\ref{xstep}), (\ref{vstep}) is consistent with the exact probability
distribution $P_g^{\rm free}(x,v;x_0,v_0;t)$ in free space, i.e. in the absence of
boundaries, of a randomly accelerated particle in a gravitational field $g=1$ with
initial conditions $x_0$, $v_0$ at $t=0$. This quantity satisfies the time-dependent
Fokker-Planck equation (\ref{timefp}) (see Appendix  A) with initial condition
(\ref{ic}). The exact form of $P_g^{\rm free}(x,v;x_0,v_0;t)$ follows immediately
from Eq. (\ref{0g}) and the expression for $P_0^{\rm free}(x,v;x_0,v_0;t)$ in Eq.
(27) of \cite{bibu}

In free space there is no time-step error in the algorithm (\ref{xstep}),
(\ref{vstep}). The time step $\Delta_{n+1}$ is arbitrary. However, for a particle on
the half line $x>0$, trajectories are not generated with the correct probability
close to the boundary. This is because $P_g^{\rm free}(x,v;x_0,v_0;t)$ includes
trajectories which travel from positive to negative $x$ and return during the time
$t$, violating the restriction $x>0$. As in Refs. \cite{bibu,kb}, we make the time
step smaller near the boundary to exclude these spurious trajectories. Generalizing
the discussion in Ref. \cite{kb} to include a gravitational field $g=1$, we find
that the spurious trajectories occur with negligible probability if the time step
satisfies
\begin{equation}
x_n+v_n\Delta_{n+1}-{1\over
2}\Delta_{n+1}^2-c\Delta_{n+1}^{3/2}>0\;,\label{criterion}
\end{equation}
where the constant $c$ is about 5 or larger.

The most efficient time step is the largest $\Delta_{n+1}$ consistent with
inequality (\ref{criterion}), which we denote by ${\cal D}(x_n,v_n)$. Using a
smaller time step slows the simulation without improving the accuracy. A convenient
choice is $\Delta_{n+1}={\cal D}(x_n,v_n)+\delta$, where a small minimum time step
has been included. Without it, the step size decreases to zero as the particle
approaches the boundary, and it never gets there.

For small $r$ the typical speed of the particle becomes extremely small after many
boundary collisions. To simulate the behavior reliably, we reduced the minimum time
step $\delta$ as the speed decreased, following Ref. \cite{kb}. After each boundary
collision we set the root-mean square velocity change $\Delta v=(2\delta)^{1/2}$,
corresponding to the minimum time step, equal to to 1/500 of the velocity just after
the collision and then used this value of $\delta$ until the next boundary
collision. This $\Delta v$ is the smallest velocity the algorithm can handle
reliably.

\subsection{Results}
According to Eq. (\ref{collrate}) the equilibrium rate of boundary collisions in
which the particle is reflected with velocity between $v$ and $v+dv$ is given by
$P(0,v)\thinspace v\thinspace dv$. We used this relation to determine $P(0,v)$ from
our simulations.

In Fig. 4 the simulation results for $r=0.75$ and $r=0.25$ are compared with the
exact asymptotic forms (\ref{asym1}), (\ref{asym2}) and (\ref{asym3})-(\ref{asym6})
for small and large $v$, respectively. Varying the proportionality constants in Eqs.
(\ref{asym1}) and (\ref{asym3}) shifts the dashed and solid curves vertically in the
figure, without changing the slope. The proportionality constants were chosen to
best fit the simulation data, and the fit is excellent.

In Fig. 5 the simulation results for $r={1\over 2}$ and $r={1\over 3}$ are compared
with the exact analytical predictions (\ref{P(0,v)}) and (\ref{constants}). No
adjustable parameters are involved. The agreement is excellent.

\section{Concluding Remarks}

In this paper we have studied the equilibrium of a particle on the half line subject
to both a random force with zero mean and a constant gravitational force which
drives the particle toward a single inelastic boundary. It is interesting to compare
with the case \cite{bfg,bk} of a randomly accelerated particle moving on the finite
line between two inelastic boundaries with no gravitational field. In both cases the
equilibrium distribution function has the same divergent behavior $P(0,v)\sim
v^{-\beta}$ for small $v$, given in Eqs. (\ref{asym1}), (\ref{asym2}). As discussed
below Eq. (\ref{rc}), this comes from a high rate of low velocity inelastic
collisions with one of the boundaries. In both cases the random force, not gravity,
is responsible for the rapid return of a low velocity particle to the boundary.

In contrast, the equilibrium behavior for high velocities is different in the two
cases. With a single boundary and a gravitational field, $P(0,v)\sim v^{-m}e^{-qv}$
for large $v$, according to Eqs. (\ref{asym3})-(\ref{asym6}). With two boundaries
and no gravitation, $P(0,v)\sim \exp[-(v/v_{\rm ch})^3]$, according to Eqs. (14) and
(16) of Ref. \cite{bk}. Thus, in the case of two boundaries high velocities are far
more strongly suppressed. This is not surprising. Since a high velocity particle
quickly travels between the two boundaries with negligible change in velocity, the
time between inelastic collisions decreases as $\vert v\vert^{-1}$ as $\vert v\vert$
increases. In the case of a single boundary and a gravitational field, on the other
hand, a particle which leaves the boundary with a high velocity travels far from the
boundary before it returns, and the mean time between the inelastic collisions
increases linearly with $\vert v\vert$.

We conclude with quantitative answers to some simple questions one might ask on
reading the title of the paper: What is the mean height $\langle x\rangle$ of the
ball above the inelastically reflecting surface, how large is the root-mean-square
fluctuation $\Delta x=\langle(x-\langle x\rangle)^2\rangle^{1/2}$, and what is the
equilibrium collision rate $I$ ? In the two cases $r={1\over 2}$ and ${1\over 3}$
where we have found the exact equilibrium distribution function $P(x,v)$, Eqs.
(\ref{collrate}), (\ref{P(0,v)}), (\ref{constants}), and (\ref{P(x)2}) imply
\begin{equation}
\begin{array}{l} r={1\over 2}:\\
r={1\over 3}:
\end{array}\quad
\begin{array}{l}
\langle x\rangle=6\;,\\
\langle x\rangle=3.89\;,\end{array}\quad
\begin{array}{l} \Delta x=10.4\;,\\
\Delta x=7.68\;,\end{array}\quad\begin{array}{l} I={1\over 3}\;,\\I={3\over 4}\;,
\end{array}\label{xandDeltax}
\end{equation}
For both values of $r$ the fluctuations in the height of the ball are larger than
the mean height. For $r={1\over 3}$, the mean height is smaller and the collision
rate is larger than for $r={1\over 2}$. The mean height is expected to decrease as
$r$ decreases but remain positive for all $0<r<1$, since $P(x)$ remains positive and
integrable. The collision rate is expected to increase as $r$ decreases and is
infinite for $r<r_c$, as discussed below Eq. (\ref{rc}). Finally, we note that these
results for $g=\Lambda=1$ in Eq. (\ref{xandDeltax}) are easily generalized to
arbitrary values of the parameters $g$ and $\Lambda$ in Eq. (\ref{eqmo}). In
accordance with footnote \cite{rescale}, the numerical values in Eq.
(\ref{xandDeltax}) for $\langle x\rangle$ and $\Delta x$ are multiplied by
$g^{-3}\Lambda^2$ and for $I$ by $g^2\Lambda^{-1}$.

\acknowledgments{TWB thanks Dieter Forster for very informative conversations about
the function $K_{i\gamma}(v)$.}
\appendix
\section{Alternate Expressions for $G_g(v,u)$}

To discuss first-passage properties, we define $P_g(x,v;x_0,v_0;t)dxdv$ as the
probability that the position and velocity of a particle, moving on the half line
$x>0$ according to Eq. (\ref{eqmo}) with $g=\Lambda=1$, have evolved from $x_0$,
$v_0$ to values between $x$ and $x+dx$, $v$ and $v+dv$ in a time $t$ without ever
reaching $x=0$. The quantity $P_g$ satisfies the time-dependent Fokker-Planck
equation \cite{fpeq}
\begin{equation} \left({\partial\over\partial t}+v{\partial\over\partial
x}-{\partial\over
\partial v}-{\partial^2\over
\partial v^2}\right)P_g(x,v;x_0,v_0;t)=0\thinspace,\label{timefp}
\end{equation}
with the initial condition
\begin{equation}
P_g(x,v;x_0,v_0;0)=\delta(x-x_0)\delta(v-v_0)\;\label{ic} \end{equation} and the
absorbing boundary condition
\begin{equation}
P_g(0,v;x_0,v_0;t)=0\;,\quad v>0\;.\label{timebc}
\end{equation}

In the absence of the gravitational field, i.e. in the case $g=0$ instead of $g=1$,
the corresponding probability distribution $P_0(x,v;x_0,v_0;t)$ satisfies the same
Fokker-Planck equation, boundary condition, and initial condition, except that the
third term $-\partial P/\partial v$ in Eq. (\ref{timefp}) is absent. This leads to
the simple relation
\begin{equation}
P_g(x,v;x_0,v_0;t)=e^{-(v-v_0)/2-t/4}\;P_0(x,v;x_0,v_0;t)\label{0g}
\end{equation}
between the distribution functions with and without gravity.

As discussed below Eq. (\ref{kernel1}), the quantity $vG_g(v,u)dv$ in integral
equation (\ref{inteq}) represents the probability that a particle moving according
to Eq. (\ref{eqmo}) with initial position $x=0$ and initial velocity $u>0$ arrives
with speed between $v$ and $v+dv$ on its first return to $x=0$. Since the return
occurs between $t$ and $t+dt$ with probability $vP_g(0,-v;0,u;t)\;dv\;dt$,
\begin{equation}
G_g(v,u)=\int_0^\infty dt\thinspace P_g(0,-v;0,u;t)=
e^{(v+u)/2}\tilde{P}_0(0,-v;0,u;\textstyle{1\over 4})\;,\label{GgP}
\end{equation}
where we have made use of Eq. (\ref{0g}) and introduced the Laplace transform
\begin{equation}
\tilde{P}_0(x,v;x_0,v_0;s)=\int_0^\infty dt\thinspace e^{-st} P_0(x,v;x_0,v_0;t)\;.
\label{mainpointofA}
\end{equation}

In a classic paper on the first passage properties of a randomly accelerated
particle, McKean \cite{mck} derived the exact form of $P_0(0,-v;0,u;t)$ and its
Laplace transform. The expressions for $G_g(v,u)$ in Eqs. (\ref{kernel2}) and
(\ref{kernel3}) follow from Eq. (\ref{GgP}) and McKean's Eqs. (5) and (6).

\section{Derivation of exact $P(0,v)$ for $r={1\over 2}$ and ${1\over 3}$}

Expression (\ref{kernel2}) for $G_g(v,u)$ and the Kontorovich-Lebedev transforms
\cite{lktransform}
\begin{eqnarray}
f(v)&=&\int_0^\infty d\gamma\;g(\gamma)K_{i\gamma}(v)\;,\label{lk1}\\
g(\gamma)&=&2\pi^{-2}\gamma\sinh(\pi\gamma)\int_0^\infty dv\thinspace
v^{-1}f(v)K_{i\gamma}(v),\label{lk2}
\end{eqnarray}
imply
\begin{equation}
v^{-1}e^{v/2}K_{i\gamma}(v)=2\cosh\left({\pi\gamma\over 3}\right)\int_0^\infty
du\thinspace G_{g}(v,u)e^{-u/2}K_{i\gamma}(u)\;.\label{binteq}
\end{equation}
For $i\gamma\to{1\over 2}$, $K_{i\gamma}(v)\to {\rm const}\times v^{-1/2}e^{-v}$,
and integral equation (\ref{binteq}) is entirely equivalent to Eq. (\ref{inteq}),
with $r={1\over 3}$ and $P(0,v)\propto v^{-3/2}e^{-3v/2}$. Thus, we have found the
exact $P(0,v)$ for $r={1\over 3}$, already announced in Eq. (\ref{P(0,v)}).

The other result in Eq. (\ref{P(0,v)}), that $P(0,v)\propto v^{-1}e^{-v}$ for
$r={1\over 2}$ is an exact solution of the integral equation (\ref{inteq}), is
readily proved by integrating Eq. (\ref{binteq}) over $\gamma$ from $\gamma=0$ to
$\infty$ and then using the results,
\begin{eqnarray}
&&\int_0^\infty d\gamma\thinspace K_{i\gamma}(v)=
{\pi\over 2}\thinspace e^{-v}\;,\label{ident1}\\
&&\int_0^\infty d\gamma\thinspace \cosh\left({\pi\gamma\over 3}\right)
K_{i\gamma}(v)={\pi\over 2}\thinspace e^{-v/2}\;.\label{ident2}
\end{eqnarray}
These relations, which are examples of the Kontorovich-Lebedev transforms
(\ref{lk1}), (\ref{lk2}), follow directly from Ref. \cite{gr}, Eq. (6.795-1).

\clearpage \noindent\Large\textbf{Figure Captions} \normalsize \vspace{1.0cm}
\begin{description}
\item{Figure 1:} The exponents $m(r)$ and $\beta(r)$, given by Eqs. (\ref{asym1}),
(\ref{asym2}), and (\ref{asym3})-(\ref{asym6}). Note the discontinuity in $m(r)$ at
$r={1\over 2}$. \item{Figure 2:} The decay constant $q(r)$, given by Eqs.
(\ref{asym4}) and (\ref{asym6}). Both $q(r)$ and $dq(r)/dr$ are continuous at
$r={1\over 2}$, but $d^2q/dr^2$ is discontinuous. \item{Figure 3:} Exact $P(x,v)$,
given by Eqs. (\ref{fpsol}), (\ref{psi}), (\ref{weightfunc2}), and
(\ref{constants}), for $r={1\over 2}$ and, from top to bottom, $x=0$, 0.001, 0.01,
and 0.1. \item{Figure 4:} Double-logarithmic plot (base 10) of $P(0,v)$ for $r=0.75$
and $0.25$. The points are the results of our simulations. The dashed and solid
lines show the predicted asymptotic forms (\ref{asym1}), (\ref{asym2}) for small $v$
and (\ref{asym3})-(\ref{asym6}) for large $v$, respectively, with proportionality
constants chosen to fit the simulation data. \item{Figure 5:} Double-logarithmic
plot (base 10) of $P(0,v)$ for $r={1\over 2}$ and ${1\over 3}$. The points are the
results of our simulations. The solid curves show the exact theoretical predictions
(\ref{P(0,v)}) and (\ref{constants}), with no adjustable parameters.
\end{description}

\clearpage
\begin{figure}[bkfig1]
\begin{center}
   \includegraphics*[width=0.7\textwidth]{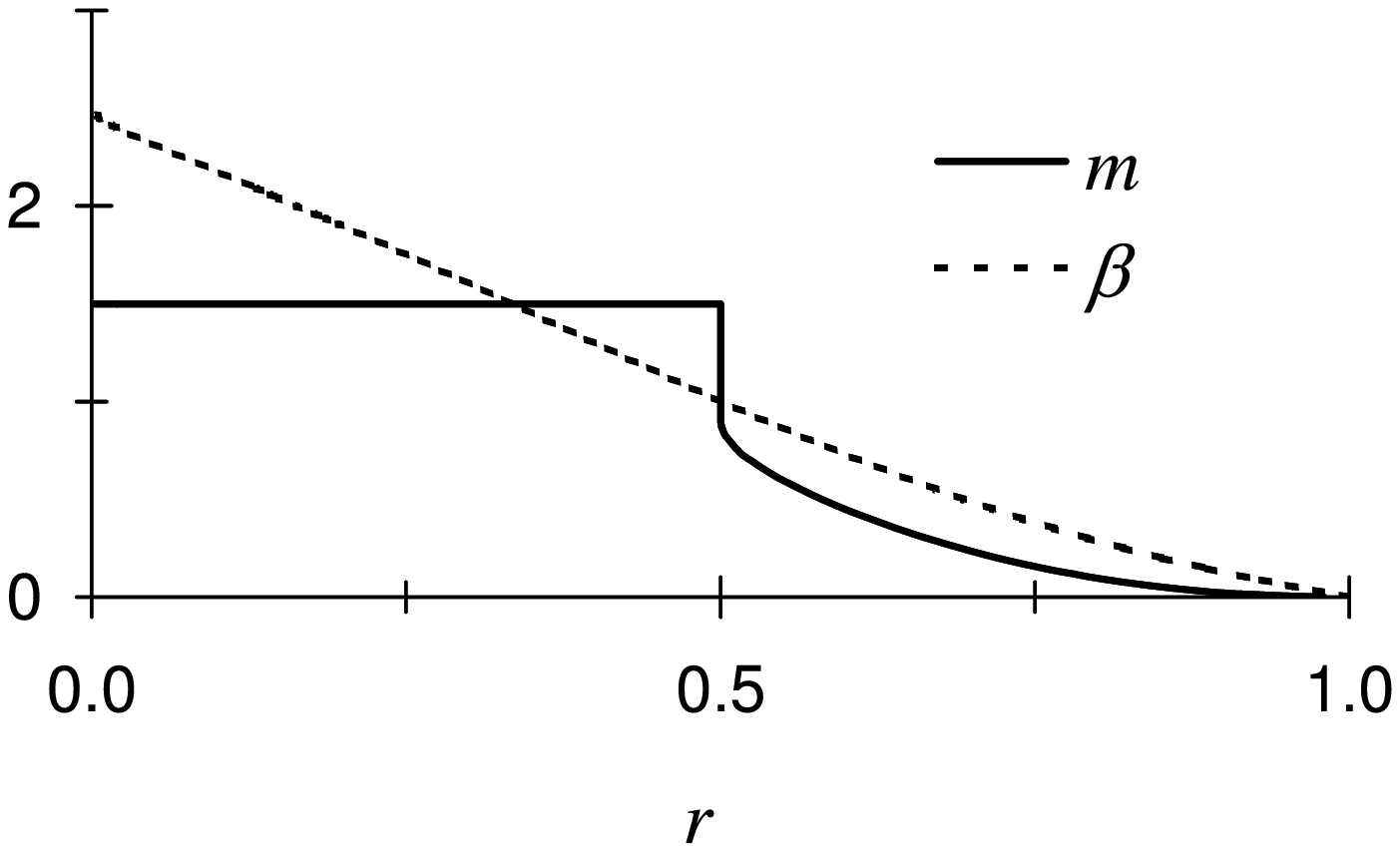}\\*
\end{center}
\caption{The exponents $m(r)$ and $\beta(r)$, given by Eqs. (\ref{asym1}),
(\ref{asym2}), and (\ref{asym3})-(\ref{asym6}). Note the discontinuity in $m(r)$ at
$r={1\over 2}$.}\label{fig1}
\end{figure}

\clearpage
\begin{figure}[bkfig2]
\begin{center}
   \includegraphics*[width=0.7\textwidth]{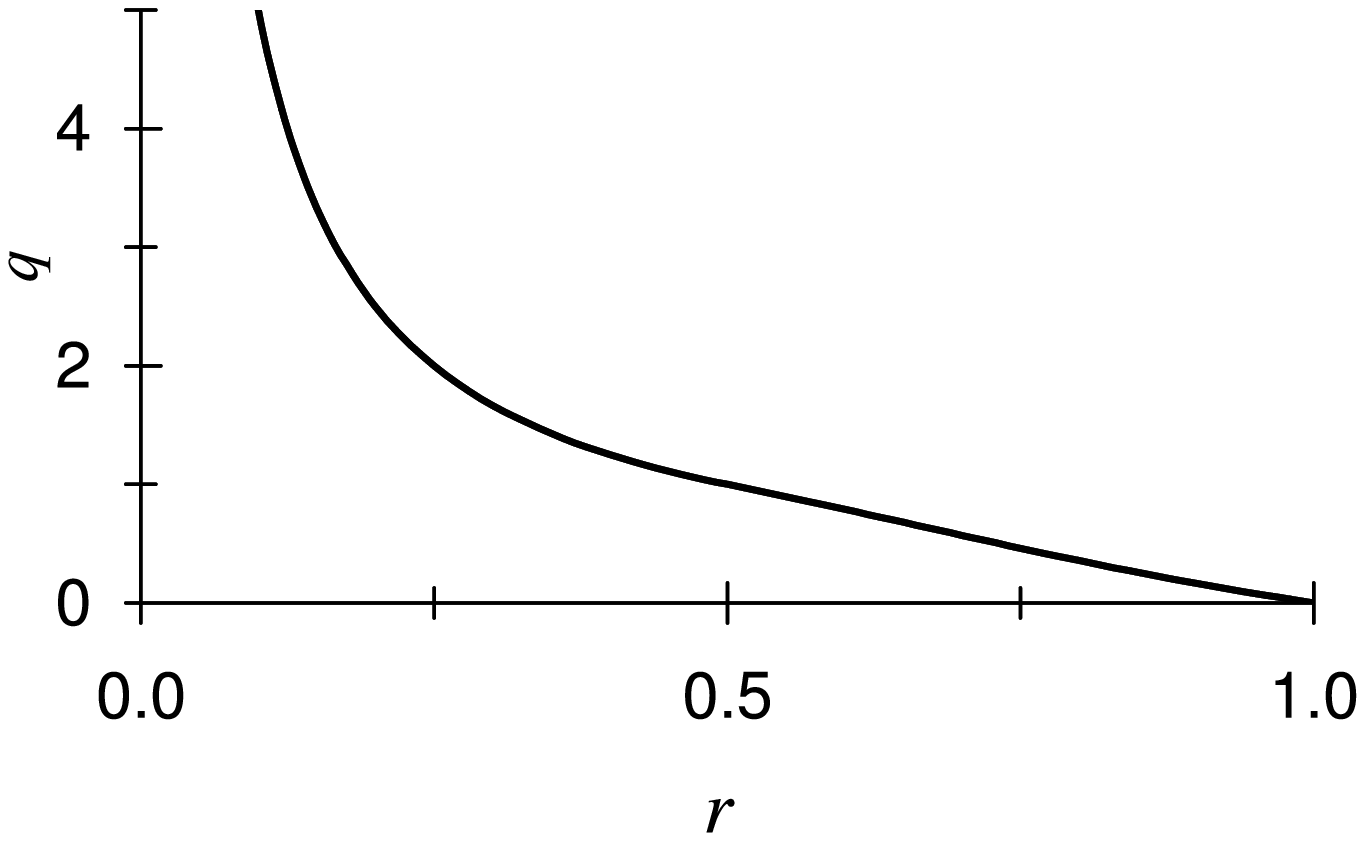}\\*
\end{center}
\caption{The decay constant $q(r)$, given by Eqs. (\ref{asym4}) and (\ref{asym6}).
Both $q(r)$ and $dq(r)/dr$ are continuous at $r={1\over 2}$, but $d^2q/dr^2$ is
discontinuous.}\label{fig2}
\end{figure}

\clearpage
\begin{figure}[bkfig3]
\begin{center}
   \includegraphics*[width=0.7\textwidth]{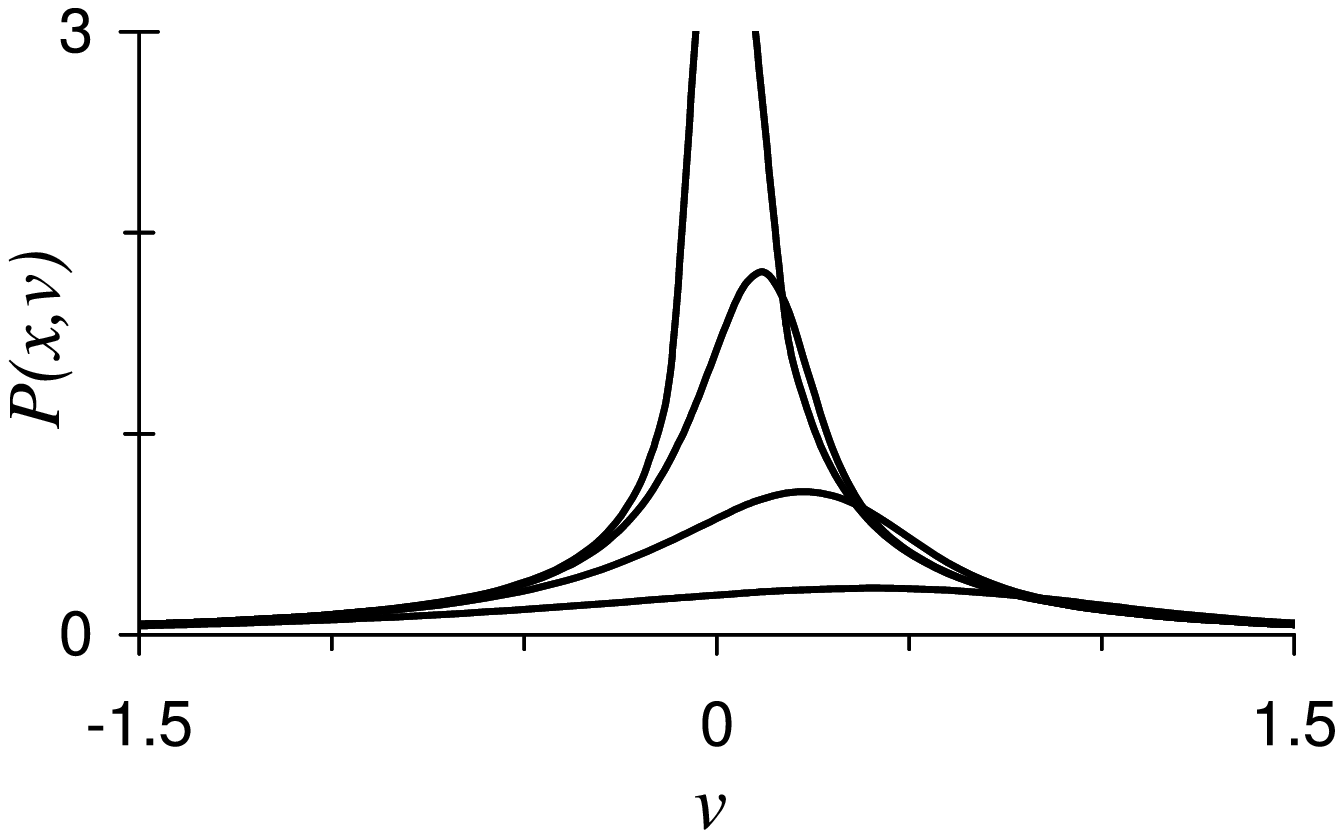}\\*
\end{center}
\caption{Exact $P(x,v)$, given by Eqs. (\ref{fpsol}), (\ref{psi}),
(\ref{weightfunc2}), and (\ref{constants}), for $r={1\over 2}$ and, from top to
bottom, $x=0$, 0.001, 0.01, and 0.1.}\label{fig3}
\end{figure}

\clearpage
\begin{figure}[bkfig4]
\begin{center}
   \includegraphics*[width=0.7\textwidth]{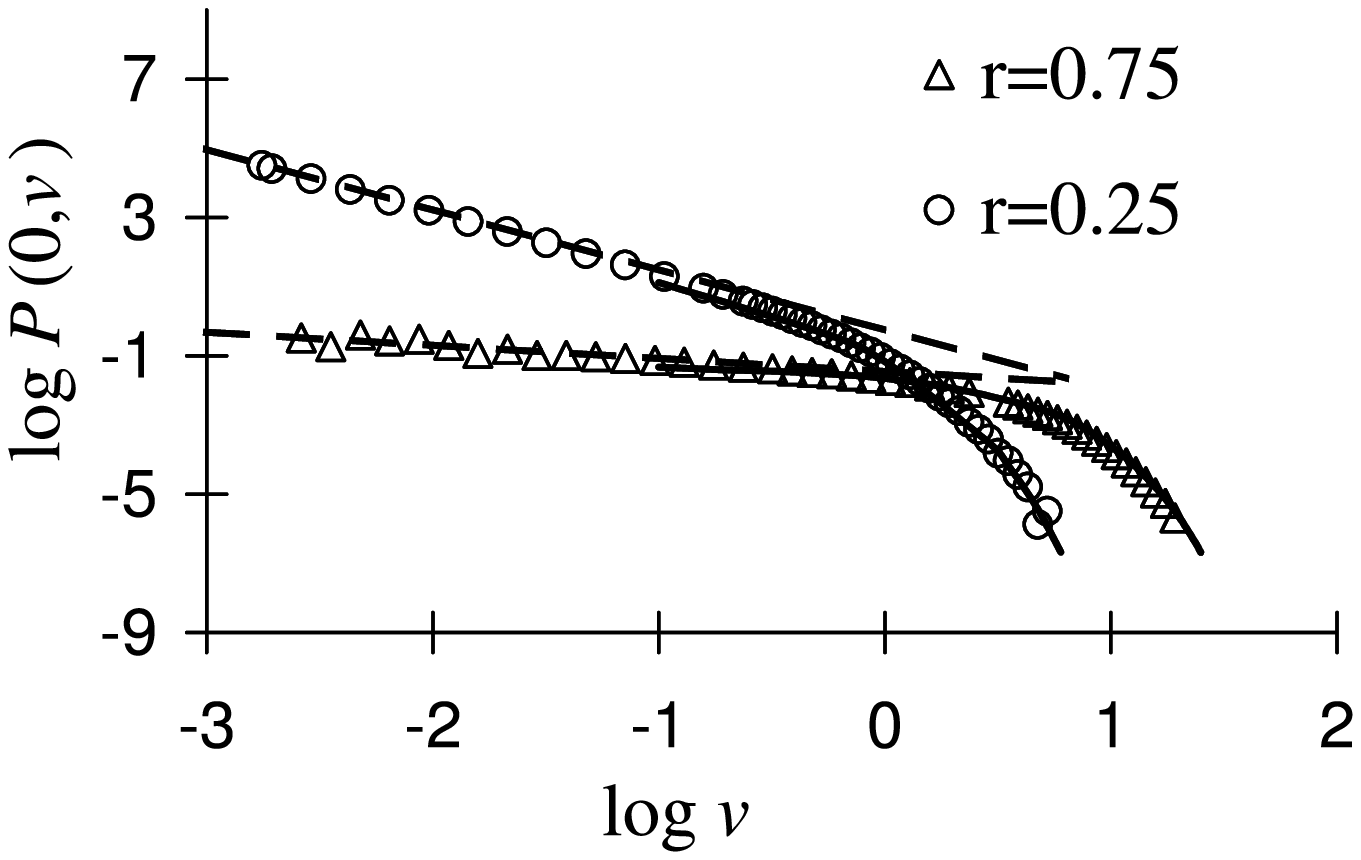}\\*
\end{center}
\caption{Double-logarithmic plot (base 10) of $P(0,v)$ for $r=0.75$ and $0.25$. The
points are the results of our simulations. The dashed and solid lines show the
predicted asymptotic forms (\ref{asym1}), (\ref{asym2}) for small $v$ and
(\ref{asym3})-(\ref{asym6}) for large $v$, respectively, with proportionality
constants chosen to fit the simulation data.}\label{fig4}
\end{figure}

\clearpage
\begin{figure}[bkfig5]
\begin{center}
   \includegraphics*[width=0.7\textwidth]{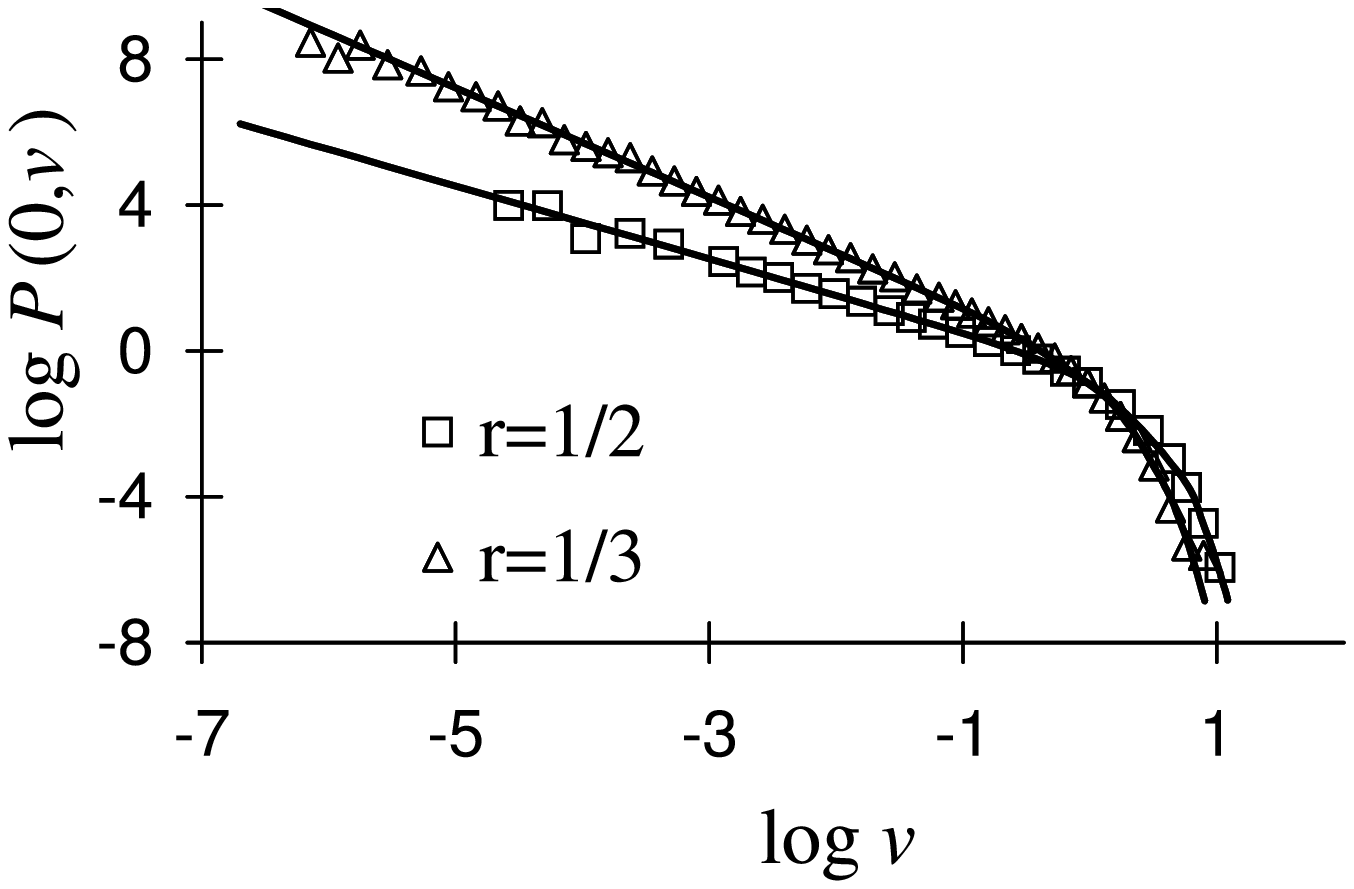}\\*
\end{center}
\caption{Double-logarithmic plot (base 10) of $P(0,v)$ for $r={1\over 2}$ and
${1\over 3}$. The points are the results of our simulations. The solid curves show
the exact theoretical predictions (\ref{P(0,v)}) and (\ref{constants}), with no
adjustable parameters.}\label{fig5}
\end{figure}

\end{document}